\documentstyle[prl, aps, twocolumn, psfig]{revtex}

\newcommand\be{\begin{equation}}
\newcommand\ee{\end{equation}}
\newcommand\bd{\begin{displaymath}}
\newcommand\ed{\end{displaymath}}
\newcommand\bea{\begin{eqnarray}}
\newcommand\eea{\end{eqnarray}}

\newcommand\lt{\left}
\newcommand\rt{\right}

\begin{document}

\twocolumn[

\hsize\textwidth\columnwidth\hsize\csname @twocolumnfalse\endcsname

\draft

\title{Long-Lived Localized Field
Configurations in Small Lattices: Application to Oscillons }

\author{M. Gleiser\footnote{email: gleiser@dartmouth.edu}}
\address{Department of Physics and Astronomy, 
Dartmouth College, Hanover, NH 03755, USA}

\author{A. Sornborger\footnote{email:ats@camelot.mssm.edu. Address after
September 7th: Laboratory for Applied Mathemtics, Mt. Sinai School of Medicine,
One Gustav L. Levy Place, New York, NY 10029.}}
\address{NASA/Fermilab Astrophysics Group, Fermi National Accelerator
Laboratory, Box 500, Batavia, IL 60510-0500, USA}

\maketitle

\begin{abstract}
Long-lived localized field configurations such as breathers, oscillons,
or more complex objects naturally arise in the context of a
wide range of nonlinear models in different numbers of spatial
dimensions. We present a numerical method, which we call the {\it
adiabatic damping method}, designed to study such configurations in
small lattices. Using 3-dimensional oscillons in $\phi^4$ 
models as an example, we show that the method accurately (to a part
in $10^5$ or better) reproduces results obtained with static
or dynamically expanding lattices, dramatically cutting down in
integration time. We further present new results for 2-dimensional
oscillons, whose lifetimes would be prohibitively long to study with
conventional methods.

\end{abstract}

\pacs{PACS numbers: 05.10.-a; 11.27.+d; 05.45.Yv; 47.11.+j}

]

\section{Introduction}

The study of long-lived, localized coherent configurations is of great interest
to many areas of physics and engineering. In general, these structures arise
within the context of numerical studies of effective nonlinear field models, 
which may either
describe behavior already observed in the laboratory or conjectured to 
exist in phenomena yet to be observed. In contrast to the usual 
solitonic behavior, which is marked by localized time-independent 
configurations, these are time-dependent configurations, which nevertheless
persist for very long times. It is reasonable to suppose that the long
lifetimes are due to energy exchange promoted by nonlinear coupling
between different modes which efficiently
suppresses the radiation of energy away
from the configuration. One may collectively call these {\it persistent
coherent field configurations}, (PCFCs) in order to distinguish them
from the usual static coherent configurations which characterize
solitonic behavior. 

Perhaps the most well-known PCFCs are the 1-dimensional breathers which appear
during low velocity kink-antikink interactions in sine-Gordon and $\phi^4$
models \cite{BREATHERS}. These are bound states characterized by nonlinear
oscillations about the energy minimum (or vacuum) of the model, 
typically one of the minima in a degenerate double-well potential. As argued
by Campbell {\it et al.}, breathers should form when the kinks have enough
time to lose energy through their interaction, adjusting to their new
trapped state. It is thus expected that these bound states form
for small relative velocities for the kink-antikink pair, although the 
dependence on the velocity is far from trivial \cite{CAMPBELL}. Once the
breather forms, it will remain in its oscillatory state for a remarkably
long time, with minimal emission of radiation. We are not aware of a detailed
study of the lifetimes of these configurations for obvious reasons: it would
take a huge amount of integration time in order to follow the evolution of
these bound states until their demise. The nonlinear nature of the problem
also has precluded (at least so far) analytical estimates for breather
lifetimes. 

In the mid-1970's, 3-dimensional, breather-like configurations were
discovered in the context of $\phi^4$ models by Bogolubsky and Makhankov
\cite{BOGOLUBSKY}. These authors found that, for certain initial conditions,
spherically-symmetric bubbles settled into a long-lived configuration which
they called pulsons. In 1994, one of us rediscovered these configurations
while studying the collapse of subcritical bubbles in the context of
degenerate and non-degenerate $\phi^4$ models \cite{GLEI1}. Since the typical
behavior characterizing these configurations is the high-frequency oscillation
about the global minimum of the model, the name {\it oscillon} was chosen
instead. A further detailed analysis revealed more of the remarkable behavior
of these configurations, such as the dependence of lifetime on initial
radius, the minimal radius for the bubbles to settle into oscillonic behavior,
and the mechanism for their final demise \cite{GLEI2}. Recent work has
pointed out the possible relevance of these structures for resonant hadronic
states \cite{HSU}.

These configurations are far from being constrained to relativistic nonlinear
field theories. Remarkably similar behavior has been found in experiments
involving grains (or ``sand'') placed on a plate undergoing sinusoidal 
vibrations \cite{SWINNEY1}. These PCFCs were independently named oscillons, 
and their discovery has triggered a host of theoretical work attempting to
model the experimental 
results. These include molecular dynamics simulations \cite{MOLDYN},
semi-continuum theories \cite{SEMICONT}, Ginzburg-Landau models \cite{GL},
coupled-map models \cite{COUPLEDMAP}, order-parameter models \cite{ORDERPARAM},
and other continuum models \cite{EGGERS}. The difficulty here is in obtaining
macroscopic laws describing the motion of granular materials, which can 
exhibit both solid and fluid properties. Oscillonic behavior has also been
found in studies of acoustic instabilities in stars\cite{SUN}.
Long-lived, spatially-extended oscillatory behavior appears to be bringing
together research in traditionally very distant fields of physics.

Our goal in this paper is to present a method which is extremely useful in the
study of PCFCs arising in the context of 
nonlinear wave models. The numerical challenge arises due to the peculiar
nature of these configurations; although they are well-localized, they do
radiate some of their energy, which, for small lattices, will get reflected
back, compromising the numerical results. Furthermore, a large amount
of energy may be shed initially as the field (or fields) settles (settle)
into a PCFC.
The simplest approach is to set up a very large
spatial lattice, large enough that the outward radiation will never
hit the boundary within the integration time. Clearly, this approach
(used in \cite{GLEI1}) is extremely inefficient for very long-lived PCFCs.
An improvement is to use dynamically expanding lattices, that is, lattices that
grow ahead of the radiation \cite{GLEI2}. This saves some time, but not much as the lifetime
becomes fairly large. Clearly, for more detailed studies of these objects,
a more efficient numerical method is badly needed, one that allows for an
accurate study of very long-lived PCFCs with relatively small lattices. 
Although we will introduce the method within the
context of $d$-dimensional spherically-symmetric models, it can be easily 
implemented in more complex situations. 

The paper is organized as follows. In the next section we present in
detail the numerical integration routine we used, the 4th-order
operator splitting method developed in Ref. \cite{ANDREWEWAN}. This is
followed by a discussion of the adiabatic damping method we propose
for studying long-lived configurations in small lattices. In section 3
we use 3-dimensional oscillons to test the accuracy of the adiabatic
damping method by comparing its performance with results obtained from
dynamically expanding lattices. In section 4 we present an analytical
estimate for the minimum radius for the onset of 2-dimensional
oscillonic behavior. This estimate is then tested numerically in
section 5, where we also display the result for the lifetime of a few
oscillons, which live at least 3 orders of magnitude longer than their
3-dimensional counterparts. We conclude in section 6 with a summary of
our results and directions for future work.

\section{Numerical Techniques}

\subsection{4th order Symplectic Method}

The Hamiltonian field equations for our theory are
\begin{equation}
  \dot\pi = \partial_\rho^2\phi
            + \frac{(d-1)}{\rho}\partial_\rho\phi
            - \partial_\phi V(\phi)
\end{equation}
and
\begin{equation}
  \dot\phi = \pi ~~,
\end{equation}
where $d$ is the number of spatial dimensions, and $V(\phi)$ is the potential,
taken to be a function of the field $\phi$.
To integrate these equations, we use a higher-order operator splitting
method, which is symplectic for Hamiltonian systems. In brief,
symplectic methods use the fact that Hamiltonian systems of equations
can be written as
\begin{equation}
  \dot z = \{z, H\}~~,
\end{equation}
where the vector $z = (\pi_i, \phi_i)$ (in our case, $i = 1$, so we
drop the $i$'s), and where $\{a, b\}$ is a Poisson bracket. Now, we
define the operator $D_H$ by $D_H z \equiv \{z, H\}$. The equations
become
\begin{equation}
  \dot z = D_H z ~~,\label{firstordereq}
\end{equation}
which, integrating for time $\Delta t$, has the formal solution
\begin{equation}
  z(t + \Delta t) = e^{\Delta t D_H} z(t)~~.
\end{equation}
Note that, in general, systems of first-order equations can also be
written in the form (\ref{firstordereq}). However, the phase-space
behavior of non-symplectic systems is usually singular. In general,
$D_H$ is a sum of terms $D_H = D_1 + D_2 + ... + D_N$. To some order
in $\Delta t$, we can approximate the exponential of a sum of
operators as a product of exponentials, where each exponential has one
of the operators as its argument. We use the notation
\begin{equation}
  \lt(\Delta t \rt)
\end{equation}
to represent
\begin{equation}
  \lt( e^{\Delta t D_1} e^{\Delta t D_2} ... e^{\Delta t D_N} \rt)
\end{equation}
and
\begin{equation}
  \lt(\Delta t \rt)^T
\end{equation}
to represent
\begin{equation}
  \lt( e^{\Delta t D_N} ... e^{\Delta t D_2}e^{\Delta t D_1} \rt)~~.
\end{equation}
So, for example, the 2nd order method
\begin{equation}
  \lt( e^{\Delta t D_1}e^{\Delta t D_2}...e^{\Delta t D_N} \rt) 
  \lt( e^{\Delta t D_N}...e^{\Delta t D_2}e^{\Delta t D_1}
  \rt)
\end{equation}
is represented by
\begin{equation}
  (\Delta t)(\Delta t)^T~~.
\end{equation}
In this paper, we use an explicit fourth-order method for `splitting'
the operator $\exp(D_H \Delta t)$. The method is \cite{ANDREWEWAN}
\begin{eqnarray}
  (\Delta t)^T(\Delta t)(\Delta t)^T(-2 \Delta t)(\Delta t)^T
  (\Delta t)^T \nonumber \\
  (\Delta t)^T(\Delta t)^T(\Delta t)(\Delta t)^T
  (\Delta t)(\Delta t) \nonumber \\
  (\Delta t)(\Delta t)(-2 \Delta t)^T(\Delta t)
  (\Delta t)^T(\Delta t)~~.
\end{eqnarray}

We split the Hamiltonian $H = \pi^2 + (\nabla\phi)^2 + V(\phi)$ into
two parts, $H_1 = (\nabla\phi)^2 + V(\phi)$ and $H_2 = \pi^2$. The
action of the operator $\exp(\Delta t D_2)$ on $z$ is
\begin{eqnarray}
  \dot\phi &=& \{\phi, H_2\} = \{\phi, H\} \nonumber \\
           &=& \pi \nonumber \\
  \dot\pi  &=& 0
\end{eqnarray}
which corresponds to integrating equation (\ref{eqphi})
\begin{equation}
  \phi^{n+1} = \phi^n + \Delta t \pi^n~~, \label{eqphi}
\end{equation}
where the superscript $n$ indicates the timestep, while leaving
the value of $\pi$ the same. Note that this integration is exact for
these equations. The action of the operator $\exp(\Delta t D_1)$ on $z$
is
\begin{eqnarray}
  \dot\phi &=& 0 \nonumber \\
  \dot\pi  &=& \{\pi, H_1\} = \{\pi, H\} \nonumber \\
           &=& (\partial_\rho^2\phi^n 
                + \frac{(d-1)}{\rho}\partial_\rho\phi^n
                - \partial_\phi V^n)
\end{eqnarray}
which corresponds to integrating equation (\ref{eqpi})
\begin{equation}
  \pi^{n+1} = \pi^n + \Delta t (\partial_\rho^2\phi^n 
                                + \frac{(d-1)}{\rho}\partial_\rho\phi^n
                                - \partial_\phi V^n) \label{eqpi}
\end{equation}
(we use fourth-order spatial differences for the derivative terms),
while leaving the value of $\phi$ the same. Higher-order methods
combine the integration of these equations in such a manner as to
cancel higher-order errors in the commutators $[D_1, D_2]$.

The standard leapfrog method is a second-order symplectic method. To
see this from our notation, we write
\begin{eqnarray}
  (\frac{1}{2}\Delta t)(\frac{1}{2}\Delta t)^T 
     &=& \lt( e^{\frac{1}{2}\Delta t D_1} e^{\frac{1}{2}\Delta t D_2} \rt)
         \lt( e^{\frac{1}{2}\Delta t D_2} e^{\frac{1}{2}\Delta t D_1} \rt)
                              \nonumber \\
     &=& \lt( e^{\frac{1}{2}\Delta t D_1} e^{\Delta t D_2}
              e^{\frac{1}{2}\Delta t D_1} \rt)~~,
\end{eqnarray}
where $D_1 \pi = \{\pi, H\}$ and $D_2 \phi = \{\phi, H\}$. Over the
course of a simulation, we have
\begin{eqnarray}
  z_{final} &=& \lt( e^{\frac{1}{2}\Delta t D_1}
       e^{\Delta t D_2}e^{\Delta t D_1}e^{\Delta t D_2}... \rt . 
               \nonumber \\
     &&  \lt . e^{\Delta t D_1}e^{\Delta t D_2}
       e^{\frac{1}{2}\Delta t D_1} \rt) z_{initial}~~.
\end{eqnarray}
Therefore, by first putting the momenta at the half-timestep, the
leapfrog method is to alternately swap the integration of the $\phi$'s
and $\pi$'s, then finally, to correct the momenta by a
half-timestep. Higher-order methods such as the fourth-order method
above require both swapping and integration with negative timesteps.

The expected error in $\phi$ and $\pi$ from our method is $R N \Delta
t^5$, where $R$ is a coefficient of order $O(1)$ given by the
particular method and $N$ is the number of timesteps in the simulation
\cite{ANDREWEWAN}. To test the above fourth-order method, we run a
simulation with $N = 650000$ timesteps, where $\Delta t = 0.01$ on an
expanding grid. Our expected error at the end of the simulation is of
order $6.5 \times 10^{-5} R$. To measure the actual accuracy of the
method, we calculate the change in energy $\delta E/E$ over the
duration of the simulation and find a value $\delta E/E = 4 \times
10^{-4}$ over $6500$ time units, or $650000$ timesteps, which is
compatible with the expected error.

\subsection{Adiabatic Damping Method: Small Lattices for Long Simulations}

In order to decrease the size of the grid, we want to absorb the
non-linear radiation propagating away from the oscillon. For a problem
involving massless radiation, we could use absorbing boundary
conditions. However, for the problem at hand, the radiation is nonlinear,
with a different dispersion relation for different wavemodes. Therefore, we 
introduce a damping term at a sufficient distance away from the oscillon
such that the evolution of the oscillon is not disturbed, while the outgoing
radiation is absorbed. This gives us a new equation of motion
\begin{equation}
  \ddot\phi + \gamma(\rho)\dot\phi - \nabla^2\phi + V'(\phi) = 0~~.
\end{equation}
The introduction of the non-zero decay term gives us a new equation
for $\pi$
\begin{eqnarray}
  \pi^{n+1} &=& \frac{1}{1+\gamma(\rho)\Delta t/2} \lt(
                \frac{\pi^n}{1-\gamma(\rho)\Delta t/2} \rt. \nonumber \\
            &&  \lt . + \Delta t (\partial_\rho^2\phi^n 
                 + \frac{(d-1)}{\rho}\partial_\rho\phi^n
                 - \partial_\phi V^n)\rt)~~.
\end{eqnarray}
Here, $\gamma$ is a function of $\rho$. We set $\gamma = 0$ for all
gridpoints less than $\rho_0$. For gridpoints greater than $\rho_0$,
we set $\gamma = \eta^2 (\rho - \rho_0)^2$, where $\eta$ is a
{\it small} constant. For the successful implementation of the method, it
is crucial that $\rho_0$ be chosen far enough away from the PCFC so as to
not interfere with its dynamics. We find that, for a typical linear scale
characterizing the PCFC of $R_0$, a safe choice is $\rho_0 \gtrsim 20
R_0$, although different situations may call for different
choices of $\rho_0$. In figure 1, we present a diagram of the relevant scales
for the partition of the spatial lattice.

\begin{figure}
\psfig{figure=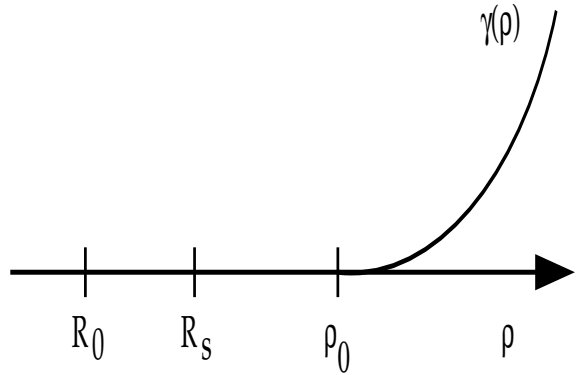,height=2.in,width=3.in}
\caption[caption]{
A schematic diagram for the relevant spatial scales in the 
implementation of the ADM. $R_0$ is the linear scale of the PCFC, $R_s$ is the
location of the shell where the energy of the PCFC is measured, and $\rho_0$
is the starting point of the damping term.
\label{damping_fig}}
\end{figure}

The introduction of the damping term reduces our method from a
symplectic method to an operator splitting method, since the phase
space of the system no longer obeys Liouville's theorem.

The smallness of $\eta$ ensures the very slow increase
of the effective damping with the radial direction. (Hence the name {\it
adiabatic damping method}.) This method can be
easily generalized for higher dimensional lattices of different geometry.
The $\rho$ dependence of $\gamma$ was chosen such that the first
derivative was zero at $\rho_0$. This choice gave us better accuracy
than, for instance, a constant or an exponential dependence on radius.

We should remark that the adiabatic damping equations are mixed
hyperbolic-parabolic. It has been claimed that \cite{SHENG}, since all
higher-order operator splitting methods have operators with backwards
evolution in time, operator-splitting methods of order greater than 2
are unstable for parabolic problems. Clearly, this is incorrect. We
were able to integrate our mixed hyperbolic-parabolic equations
for extremely long times (over $10^9$ timesteps) and
found no instability. It is also clear why these methods are
stable. The total backwards time integration in our method is $4\Delta
t$, but the method integrates forward in time for $16\Delta
t$. Therefore, any instability arising from exponential increase in
mode amplitudes from backward-time integration will be resuppressed by
the forward-time integration.

\section{Testing the Adiabatic Damping Method with 3d Oscillons}

The Lagrangian for our field
theory is
\begin{equation}
  L = \pi \int [2r]^{(d-1)} dr \lt[ \frac{1}{2} \dot\phi^2 - \frac{1}{2} \phi'^2
    - V(\phi) \rt]~~,
\label{lagrangian}
\end{equation}
where $d$ is the number of spatial dimensions (for us $d=2$ or $d=3$ only), and
we use the degenerate double-well potential,
\begin{equation}
  V(\phi) = \frac{\lambda}{4}(\phi^2 - \phi_{vac}^2)^2~~.
\label{potential}
\end{equation}

As shown in \cite{GLEI1,GLEI2}, oscillons can easily be found by setting
the initial field configuration with a Gaussian or a Tanh profile. Here, we
will use the Gaussian {\it ansatz},
\begin{equation}
  \phi = (\phi_c - \phi_{vac}) e^{-\frac{\rho^2}{R_0^2}} + \phi_{vac}~~,
\end{equation}
where $R_0$ is the core radius and $\phi_{vac}$ is the asymptotic
vacuum value of the field. $\phi_c$ sets the offset of the field from
the vacuum at the core, the central displacement from equilibrium.

The evolution of the configuration can be divided into three stages. Initially,
the field sheds enough energy to settle (or not, if the initial parameters
$R_0$ and $\phi_c$ are outside the allowed range for oscillonic behavior) into
the oscillon configuration. The lifetime of the oscillon stage is sensitive to
the choices of $R_0$ and $\phi_c$, although the energy of the configuration
is not. This is, in fact, what justifies identifying these PCFCs as a single
configuration. We conjecture that the lifetime of the oscillon configuration can
be traced to the perturbations induced by the different choices of initial
parameters, which will tend to increase the amount of radiation being emitted.
However, we so far have not been able to prove this. The final stage is the
oscillon's demise. As shown in Ref. \cite{GLEI1}, due to the small but steady
radiation from the oscillon, at some point the maximum amplitude allowed
(that is, when $\dot \phi_c =0$) falls
approximately below the inflection point of the potential, the motion
becomes linear, and $\phi_c \rightarrow \phi_{vac}$ exponentially
fast.

As one would expect, one must understand something about the evolution
of the oscillon to correctly set $\rho_0$, since damping must not interfere
with the evolution of the oscillon. As a rule of thumb, the better localized
the configuration, the closer $\rho_0$ can be to $R_0$. The fact that
oscillons behave asymptotically as $\exp(-\rho)$ certainly helps.
As remarked above, 
we found that $\rho_0 = 20 R_0$
worked well.

To make sure that the damping region we have introduced does not
interfere with the evolution of the oscillon, we checked that the
oscillon lifetime is not significantly different on grids with
damping. We simulate oscillons on two grids. One simulation is
performed on an expanding grid ($\Delta x = 0.2$), the other on a
$1024$ point grid with physical size $204.8$ ($\Delta x = 0.2$), where the
damping with constant $\eta= 0.005$ begins at $\rho = 20 R_0$. In
figure 2,
we have superimposed the graph of an expanding grid
simulation, which we take to be the exact result, and the approximate
result, which comes from a simulation with a damping region for several choices
of initial radii, $R_0$. (All physical quantities are quoted in 
dimensionless units. For 3-dimensional
relativistic field theories, distance and time are 
given in units of $\lambda^{-1/2}
\phi_{vac}^{-1}$.)
The plots
denote the time evolution of the total energy in a shell about the core
(we have set our shell at $R_s = 5 R_0$).

As is apparent from figure 2, oscillon lifetimes are essentially
indistinguishable for the time spacings of the data output in these
graphs. Evolution of oscillons generated from initial
configurations with larger radii
agrees less well with the exact oscillons due to a greater interaction
between long-wavelength modes, generated as the oscillon
decays at the end of its lifetime, and the reflecting boundary at the
far end of the damped part of the grid. 
This can be cured by using a larger grid, and increasing
$\rho_0$. Notwithstanding these considerations, the agreement in
lifetimes is still extremely good for $1024$ 
gridpoints, as is shown next. The reason that long-wavelength modes introduce
difficulties with this method is that they are damped less efficiently
than short-wavelength modes (the damping is roughly proportional to
$exp(-k^2 t)$).

\begin{figure}
\psfig{figure=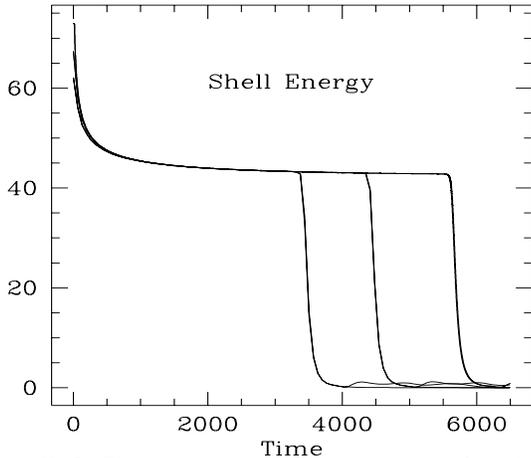,height=2.5in,width=3.0in}
\caption[caption]{Shell energy vs. Time: The evolution of the energy
in a shell of radius $R_s = 5 R_0$ about the oscillon core as a function
of time. Two graphs are superimposed, the exact solution and the
solution using a small $1024$ point grid with damping. From left to right,
we see the curves for $R_0=2.5,~2.6,~{\rm and}~2.7$.
\label{enout2.4}}
\end{figure}

In figure 3, we plot the logarithm (base ten) of the
absolute value of $\Delta E_s/E_s$ as a function of time, where $E_s$
is the energy in a shell of radius $5 R_0$ about the origin. We
chose $R_0=2.7$ in this example. $\Delta
E_s$ is the difference between the energies of the expanding grid
simulation and the damped simulation. The shell energies agree with
each other to better than one hundredth of a percent until the
oscillon decays.

\begin{figure}
\psfig{figure=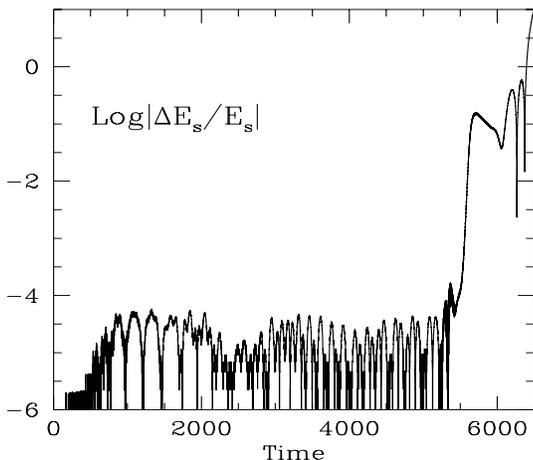,height=2.5in,width=3.0in}
\caption[caption]{$\Delta E_s/E_s$ vs. Time: The relative difference
between the expanding grid and damped grid solutions as a function of
time for $R_0 = 2.7$. Note that through all of the lifetime of the
oscillon the solutions agree to better than one hundredth of a percent.
\label{enout2.7diff}}
\end{figure}

\section{Analytical Estimate for Onset of 2-Dimensional Oscillon}

In order to estimate the approximate minimum size at which oscillon-like 
PCFCs appear in a 2-dimensional
nonlinear field theory, we begin with the {\it ansatz}
\begin{equation}
  \phi(r,t) = A(t) e^{-\frac{r^2}{R_0^2}} + \phi_{vac}~~,
\label{gaussian}
\end{equation}
where $A(t) \equiv\phi_c(t) - \phi_{vac}$. The Lagrangian for our field
theory is given in (\ref{lagrangian}) with $d=2$, and the potential is
given in (\ref{potential}) as in the case for 3d oscillons. This
{\it ansatz} restricts the time-dependence to the amplitude at the
core of the configuration, taking the radius to be a constant
parameter. Despite its simplicity, reducing a field theory to a model
with one degree of freedom, it was used successfully for determining the
minimum radius of 3d oscillons in Ref. \cite{GLEI2}.

Substituting the {\it ansatz} (\ref{gaussian}) into the Lagrangian and
integrating over the radial dimension gives an effective Lagrangian
for the single degree of freedom $A(t)$,
\begin{equation}
  L = 2\pi \lt[ \frac{R_0^2}{8} \dot A^2 - \frac{1}{4} A^2 -
    \frac{R_0^2}{32} A^4 + \frac{R_0^2}{6} A^3 - \frac{R_0^2}{4} A^2
    \rt]~~.
\end{equation}
The equation of motion is 
\begin{equation}
  \ddot A + \frac{2}{R_0^2} A + 2 A + \frac{1}{2} A^3 - 2 A^2 = 0~~.
\end{equation}
We then assume that we can separate $A(t) = A_0(t) + \delta A(t)$ and
investigate harmonic perturbations about the background (assumed stable)
solution $A_0(t)$. This leads to a frequency response of
\begin{equation}
  \omega^2(A_0, R_0) = 2 + \frac{2}{R_0^2} + \frac{3}{2} A_0^2 - 4A_0~~.
\end{equation}
If we now minimize the frequency as a function of $A_0$, we find
\begin{equation}
  \omega^2(A_0^{{\rm min}}, R_0) = \frac{2}{R_0^2} - \frac{2}{3}~~.
\end{equation}
We expect an instability to develop for $\omega^2 < 0$, giving a
bifurcation at $R_0 = \sqrt{3}$: oscillons can only exist in the presence
of these instabilities, which ensure the (temporary) survival of the nonlinear
regime. This is reminiscent of the well-known spinodal instability
in Ginzburg-Landau systems, where the growth of instabilities occurs as the
system probes the concave part of the potential (or free-energy density)
\cite{SPINODAL}. In short, although the perturbed solution is
exponentially increasing for this linear analysis, we expect nonlinear
terms to stabilize it, at least temporarily, leading to oscillon
solutions. As we will see below, this prediction is quite accurate.

\section{Results for Two-Dimensional Oscillons}

\begin{figure}
\psfig{figure=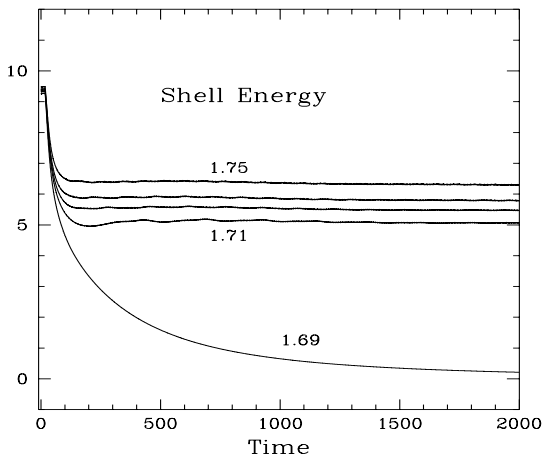,height=2.5in,width=3.0in}
\caption[caption]{Shell energy vs. time: The evolution of the energy
in a shell of radius $R_s = 5 R_0$ about the oscillon core as a function
of time. Initial
conditions were, from bottom to top, $R_0=1.69,~1.71,~1.72,~1.73,~{\rm and}~
1.75$. 
\label{min2dosc}}
\end{figure}

We can now use the adiabatic damping method to study 2d oscillons, which
preliminary studies (with large lattices) have shown to be remarkably 
long-lived. (Previous simulations were stopped as lifetimes went 
over $\tau \sim 10^4$
time units.)
In figure \ref{min2dosc}, we plot the energy in a shell of radius
$R_s = 5 R_0$ about the origin as a function of time. Starting with Gaussian
initial configurations, we searched for
oscillons around $R_0 = \sqrt{3}$, the expected onset value of the
oscillon solution obtained above, 
and found the bifurcation at the slightly lower value
of $R_0 \simeq 1.71$, as indicated in the figure.

\begin{figure}
\psfig{figure=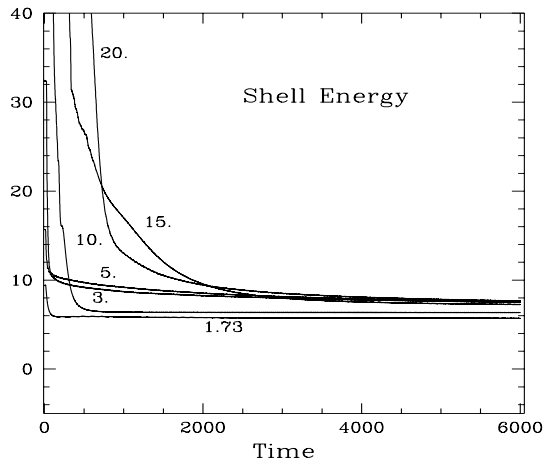,height=2.5in,width=3.0in}
\caption[caption]{Shell energy vs. time: Examples of several 2d oscillons.
The radii for the initial Gaussian profiles 
are denoted by each respective curve.
\label{shell_energy2d}}
\end{figure}

\begin{figure}
\psfig{figure=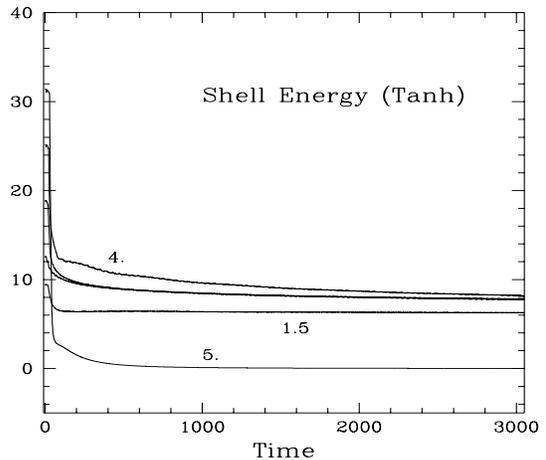,height=2.5in,width=3.0in}
\caption[caption]{Shell energy vs. time: Examples of several 2d oscillons
from a Tanh initial configuration.
The initial radii are denoted by each respective curve, with $R_0=2,{\rm and}~3
$ omitted for clarity.
\label{shell_energy2dth}}
\end{figure}

Above this value for the initial radius, our results exhibit the same behavior 
as for the 3d oscillons, with a few new and surprising properties,
which are easily identified in figure \ref{shell_energy2d}. First, we could
not find a maximum radius above which the initial configurations do not become
oscillons. (We searched all the way to $R_0=30$.) We recall that 
for 3d oscillons
obtained from Gaussian initial configurations, the maximum
radius was $R_0\simeq 4.2$ \cite{GLEI2}. We suspect this peculiar
behavior to be a consequence
of the fact that, in 2d, Gaussian configurations have constant surface energy
densities, that is, no radial dependence. [For 2d Gaussians, the energy 
goes as $E=A+BR_0^2$, where $A$ and $B$ depend on the potential.]
For initial configurations with
Tanh profiles, we found oscillons within the interval $1.5 \lesssim R_0
\lesssim 4.5$, as indicated in figure \ref{shell_energy2dth}. We note that
Tanh configurations with small radii are fairly well approximated by Gaussian
configurations, that is, have small surface energy terms.

Second, the energy plateau is not as clearly defined as in the 3d case.
For example, a configuration with initial radius $R_0=10$ evolves into
a lower energy oscillon than the other cases in the figure. Also, the
oscillon energy for the marginal case $R_0=1.73$ is lower than the rest.
This suggests that there may exist several ``excited'', or metastable,
oscillonic states
in 2-dimensional models, something worth pursuing.

Third, we also could not find a finite lifetime for the oscillons. Using
the adiabatic damping method, we followed the evolution of the oscillons for
over $10^7$ time units without observing their demise. [With our time steps
this implies over $2\times 10^9$ time-iterations.] This strongly
suggests that these may be time-dependent stable solutions. We are currently
pursuing this issue in more detail, using both analytical and numerical
methods. We note that in theories with conserved particle number,
simple time-dependent solutions known as non-topological solitons have been
found \cite{NTS}. However, in the case at hand, the conserved quantity giving 
rise to the (possible) stability of the configuration is not immediately 
obvious.

\section{Summary and Outlook}

We have presented a method designed for the numerical investigation of
long-lived field configurations such as breathers, oscillons and other
spatially-extended persistent coherent field configurations
(PCFCs). The method uses an adiabatically increasing damping term in
the equation of motion, placed safely away from the PCFC so as not to
interfere with its dynamics. We argued that, with this method,
it is possible to follow
the dynamical evolution of these objects for extremely long times, allowing us
to obtain accurate results with very small lattices. Using
3-dimensional oscillons as an example, we showed that the method
allows for accuracies of a fraction of a percent in the measurement of
physical quantities, such as the energy and lifetime,
typically of the order of a part in $10^5$. We
then applied this approach to investigate 2-dimensional
oscillons. After obtaining the minimum radius that allows for their
existence, we discovered that lifetimes can exceed $10^7$ time units. Although
not yet proven, it is possible that these time-dependent field
configurations are absolutely stable.

There are quite a few obvious avenues for future work, in addition to the ones
mentioned above. It would be straightforward to apply the method to
more complex geometries, for example searching for tube-like oscillons,
or the study of interactions between oscillons, similar to the 1-dimensional
kink-antikink scattering studies. As mentioned elsewhere, these configurations
may be thermally nucleated during phase transitions, leading to important
corrections to decay rates and completion times \cite{GLEI3}. Given the 
longevity of 2d oscillons, this may be particularly relevant to systems
in the 2-dimensional Ising universality class.
It would also be interesting to
study ``excited oscillons'', that is, non spherically-symmetric
configurations which can be expanded into a series of harmonics. Do these
configurations decay into their ground state (a normal, $\ell=0$ oscillon)
or are they completely unstable? Hopefully, future research will establish the
connection between the various oscillons described in the literature as
obeying some simple general principles. 

\section*{Acknowledgements}

We thank Ethan Honda for useful discussions. M.G. was supported in part
by an NSF Presidential Faculty Fellows award PHY-9453431. A.S. was supported
by the DOE and NASA grant NAG 5-7092 at Fermilab. M.G. thanks both the
Nasa/Fermilab Astrophysics Center and the High Energy Group at Boston 
University for their hospitality and support
during part of this work. A.S. thanks
the Department of Physics and Astronomy at Dartmouth College for its 
hospitality and support during part of this work.

\noindent
After Sept. 7, A. S. will be at: Laboratory for Applied Mathematics,
Mt. Sinai School of Medicine, One Gustave L. Levy Place, New York, NY
10029

\vspace{2 mm}
\noindent
Marcelo Gleiser: marcelo.gleiser@dartmouth.edu

\noindent
Andrew Sornborger: ats@camelot.mssm.edu

\end{document}